\documentclass[prl,a4paper,twocolumn,showpacs,superscriptaddress]{revtex4}
\usepackage{amsmath}
\usepackage{amsfonts}
\usepackage{amssymb}
\usepackage{bm}
\usepackage{graphicx}
\bmdefine\bomega{\omega}
\bmdefine\bOmega{\Omega}
\bmdefine\bnabla{\nabla}
\bmdefine\bkappa{\kappa}
\bmdefine\bphi{\phi}
\begin{document}

%\title{Vortex Loop Injection in Applied Flow and the Precursor Mechanism of Superfluid Turbulence}%
%\title{The Dynamics of Vortex Generation in Applied Flow and Onset of Superfluid Turbulence}%
%\title{Vortex Generation in Applied Flow and the Precursor Mechanism to Superfluid Turbulence}%
\title{Vortex Multiplication in Applied Flow: a Precursor to Superfluid Turbulence}

\author{A.P.~Finne}
\affiliation{Low Temperature Laboratory, Helsinki University of
Technology, Espoo, Finland }

\author{V.B.~Eltsov}
\affiliation{Low Temperature Laboratory, Helsinki University of
Technology, Espoo, Finland } \affiliation{Kapitza Institute for
Physical Problems, Moscow, Russia}

\author{G.~Eska}
\affiliation{Physikalisches Institut, Universit\"{a}t Bayreuth,
Bayreuth, Germany}

\author{R.~H\"anninen}
\affiliation{Department of Physics, Osaka City University, Osaka,
Japan}

\author{J.~Kopu}
\affiliation{Low Temperature Laboratory, Helsinki University of
Technology, Espoo, Finland }

\author{M.~Krusius}
\affiliation{Low Temperature Laboratory, Helsinki University of
Technology, Espoo, Finland }

\author{E.V. Thuneberg}
\affiliation{Department of Physical Sciences, University of Oulu,
Oulu, Finland}
\author{M.~Tsubota}
\affiliation{Department of Physics, Osaka City University, Osaka,
Japan}

\date{\today}

\begin{abstract}
A surface-mediated process is identified in $^3$He-B which
generates vortices at roughly constant rate. It precedes a faster
form of turbulence where inter-vortex interactions dominate. This
precursor becomes observable when vortex loops are introduced in
low-velocity rotating flow at sufficiently low mutual friction
dissipation at temperatures below $0.5\,T_{\rm c}$. Our
measurements indicate that the formation of new loops is
associated with a single vortex interacting in the applied flow
with the sample boundary. Numerical calculations show that the
single-vortex instability arises when a helical Kelvin wave
expands from a reconnection kink at the wall and then intersects
again with the wall.
%waves are excited by a wall reconnection kink, expand in the
%applied flow, and finally intersect again with the wall.
\end{abstract}

 \maketitle

\newcommand{\dvec}{{\hat {\bf d}}}
\newcommand{\lvec}{{\hat {\bf l}}}
\newcommand{\mvec}{{\hat {\bf m}}}
\newcommand{\nvec}{{\hat {\bf n}}}
\newcommand{\vsA}{{\bf v}_{\rm sA}}
\newcommand{\vsB}{{\bf v}_{\rm sB}}
\newcommand{\vn}{{\bf v}_{\rm n}}

%%%%%%%%%%%%%%%%%%%%%%%%%%%%%%%%%%%%%%%%%%%%%%%%%%%%%%%%%%%%%%%%%%%%%%%%%%%%
In superfluid $^3$He-B a hydrodynamic transition takes place below
$0.6\,T_{\rm c}$ from regular (laminar) vortex flow at high
temperatures to turbulent flow at low temperatures
\cite{Experiment}. The transition occurs as a function of damping
in vortex motion, the mutual friction dissipation, which increases
roughly exponentially \cite{Bevan} with temperature. If a bundle
of vortex loops is injected in applied flow, below $0.6\,T_{\rm
c}$ they interact generating new loops in a rapidly growing tangle
of vortices. The onset temperature of this turbulence depends on
the number of injected loops. At low enough temperature even a
single injected vortex ring leads to turbulence. This is
surprising since turbulence is thought to result from the
collective interaction of many loops. Also in numerical
calculations one ring does not lead to turbulence when placed in
uniform applied bulk flow. What is the explanation of this
conflict?

Liquid helium flow is generally contained inside solid walls and
thus the interaction of the expanding vortex ring with the sample
boundary has to be examined. Our measurements indicate that
initially the injected ring generates new vortices while
interacting with the container wall. This is observed as slow
vortex formation which precedes the later more rapid turbulence
and has a lower onset temperature. It supplies new vortices so
that ultimately rapid turbulence will switch on at some location
where the loop density has grown sufficiently.

The existence of a slow precursor, which later escalates to rapid
turbulence, is duplicated in numerical simulations. They show that
a helical Kelvin wave \cite{Donnelly,Glaberson} expands on a
single vortex when it becomes aligned sufficiently parallel to the
flow. A growing Kelvin wave may then reconnect at the boundary,
creating one new vortex as well as sharp kinks. These kinks excite
new Kelvin waves \cite{Svistunov}, starting a self-repeating
process of vortex multiplication. We call multiplication a process
in which new vortices are formed as a result of the dynamic
evolution of existing vortices.

%%%%%%%%%%%%%%%%%%%%%%%%%%%%%%%%%%%%%%%%%%%%%%%%%%%%%%%%%%%%%%%%%%%%%%%%%%%%

{\it Experiment:}---We measure with NMR the evolution in the
number of vortices when vortex loops are introduced with different
techniques in applied flow. Our sample of $^3$He-B is contained in
a quartz tube of length $d = 110\,$mm and inner radius $R =
3\,$mm. The flow is created by rotating the sample around its axis
with constant angular velocity $\bm{\Omega}$. Initially the sample
is vortex-free and the applied flow arises from the difference of
the normal and superfluid velocities, the counterflow velocity
$\bm{v}_{\rm cf}=\bm{v}_{\rm n} -\bm{v}_{\rm s}$. If viewed from
the laboratory, the normal component is in solid body rotation,
$\bm{v}_{\rm n} = \bm{\Omega} \times \bm{r}$, while the superfluid
component is stationary, $\bm{v}_{\rm s} = 0$. The maximum
velocity $v_{\rm cfm}= \Omega R$ is at the cylindrical boundary.
Such a state is possible if $v_{\rm cfm}$ is maintained below a
container-dependent critical value \cite{Parts}. This requirement
is here observed so that the growth in vortex number $N(t)$
monitors vortex multiplication after injection.

In the over-damped regime of vortex motion $T > 0.6\,T_{\rm c}$,
the injected loops expand to rectilinear lines, conserving their
number. In their lowest energy state they form a central cluster
where the $N$ straight lines are packed with an areal density
$n=2\Omega/\kappa$. Here $\kappa=h/2m_3$ is the circulation
quantum.  Outside the cluster the counterflow increases from zero
to $v_{\rm cfm} = [1-N/(\pi R^2n)]\Omega R$ at the cylindrical
boundary. Well within the under-damped regime $T < 0.6\,T_{\rm
c}$, the injection ultimately always leads to rapid turbulence.
Its signature is to create close to the equilibrium number of
lines, $N \lesssim \pi R^2 (2\Omega/\kappa)$. $N$ can be deduced
at the top and bottom of the long sample by measuring the NMR
spectra with two independent spectrometers \cite{Experiment}. The
technique is based on the strong dependence of the spectrum on
$v_{\rm cf}$. A calibration can be constructed experimentally or
by calculating the order parameter texture and its NMR spectrum
numerically \cite{Kopu}.

In the onset regime $T \lesssim 0.6\,T_{\rm c}$, the evolution
after injection depends on the injection method. The highest onset
for rapid turbulence is measured by making use of the properties
of the AB interface between a short section of magnetic-field
stabilized $^3$He-A and the remaining $^3$He-B in the long sample
\cite{Experiment}. The AB interface undergoes an instability when
$\Omega$ is increased across a well-defined critical value
$\Omega_{\rm cAB} = 1.2$ -- $1.6\,$rad/s and a bundle of
 $\sim 10$ closely spaced loops is tossed on the B-phase
side. Measured in this way, the onset is at $0.59\,T_{\rm c}$ (at
29.0\,bar pressure). Its half width of $0.03\,T_{\rm c}$ we
interpret to reflect the variation in the number of injected
loops.

The lowest onset is measured by exploiting vortex formation following
absorption of a thermal neutron in $^3$He-B \cite{Experiment}. This
reaction heats a blob of $\sim 100\,\mu$m diameter to the normal state.
While the blob rapidly cools back to the ambient temperature a vortex ring
with a diameter similar to the blob size is created. The ring starts to
inflate and evolve, if $\Omega$ exceeds a
critical value $\Omega_{\rm cn} = 1.4\,$rad/s. At
$0.59\,T_{\rm c}$ a single ring does not lead to turbulence, but at
$0.45\,T_{\rm c}$ well above 80\,\% of the neutron absorption events at
$\Omega = 1.6\,$rad/s develop to turbulence.

This comparison shows that an additional mechanism (requiring a
lower value of damping) is needed to start turbulence from a
single injected ring than when a bundle of many loops is used. We
assume that in the latter case rapid turbulence is switched on
immediately at the injection site, while in the former case a
precursor mechanism is required. In both cases at $\Omega \sim
1.4\,$rad/s the flow velocity is so high that rapid turbulence
follows within seconds and no time is left to monitor a precursor.

%%%%%%%%%%%%%%%%%%%%%%%%%%%%%%%%%%%%%%%%%%%%%%%%%%%%%%%%%
\begin{figure}[tb]
\centerline{\includegraphics[width=0.8\linewidth]{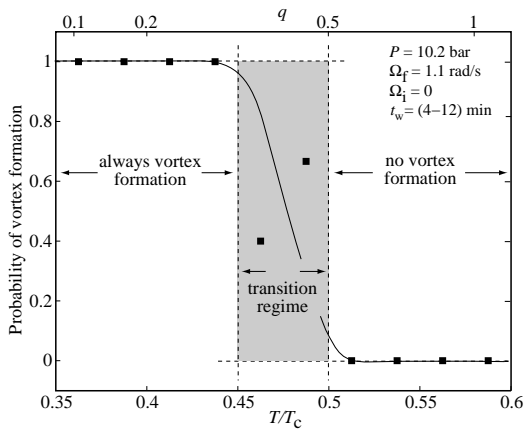}}
\caption{The probability that vortex multiplication will start
from a remnant vortex, when rotation is increased from
$\Omega_{\rm i}$ to $\Omega_{\rm f}$, plotted vs. temperature
(bottom axis) and mutual friction parameter $q = \alpha
/(1-\alpha^{\prime})$ (top axis). No vortices are formed above
$0.5\,T_{\rm c}$. Below $0.45\,T_{\rm c}$ vortex formation always
leads to turbulence. At intermediate temperatures only part of the
runs result in vortex formation.}
\label{SlowVorFormTempDependence}
\end{figure}
%%%%%%%%%%%%%%%%%%%%%%%%%%%%%%%%%%%%%%%%%%%%%%%%%%%%%%%%

These two injection techniques can be accurately controlled, but
here the flow velocity cannot be reduced to arbitrary low values.
For this two new methods were developed. Conceptually simplest is
the case of the remnant vortex at $\Omega = 0$ \cite{remnant}.
When rotation is stopped, vortices are rapidly pushed owing to
their mutual repulsion to the boundaries for annihilation. Only
the annihilation time of the last one or two vortices becomes long
below $0.5\,T_{\rm c}$, owing to the small dissipation $\alpha
(T)$ and the long sample length. A single straight vortex parallel
to the cylinder axis at a distance $b = \delta R$ from the center
survives for a time $[2\pi R^2/(\alpha \kappa)] \, [-\ln{\delta} -
\frac{1}{2} (1-\delta^2)]$, where the prefactor equals 1\,h at $
0.4\, T_{\rm c}$. In practice the annihilation time can be longer,
since the sample and rotation axes cannot be aligned perfectly
parallel and the last vortex is not straight. If the time $t_{\rm
w}$ spent at $\Omega = 0$ is shorter, then the sample contains a
remnant vortex of complex shape.

In Fig.~\ref{SlowVorFormTempDependence} we plot the probability of
observing rapid turbulence after the rotation has been increased to
$\Omega_{\rm f} = 1.1\,$rad/s with a remnant vortex in the sample. The
abrupt temperature dependence in this figure is characteristic of a
transition as a function of rapidly varying mutual friction. No
conventional explanation in terms of critical velocities or vortex mill
behavior is capable of producing such a steep jump. It is centered at
$0.47\,T_{\rm c}$ while the corresponding value with injection from the AB
interface is at $0.52\,T_{\rm c}$ (at 10.2\,bar). In both cases the half
widths are $0.03\,T_{\rm c}$.  We interpret that these two transitions do
not overlap and that the case of the remnant vortex in
Fig.~\ref{SlowVorFormTempDependence} exemplifies the transition in the
single-vortex regime where a precursor is required.

%%%%%%%%%%%%%%%%%%%%%%%%%%%%%%%%%%%%%%%%%%%%%%%%%%%%%%%%%
\begin{figure}[t]
\centerline{\includegraphics[width=\linewidth]{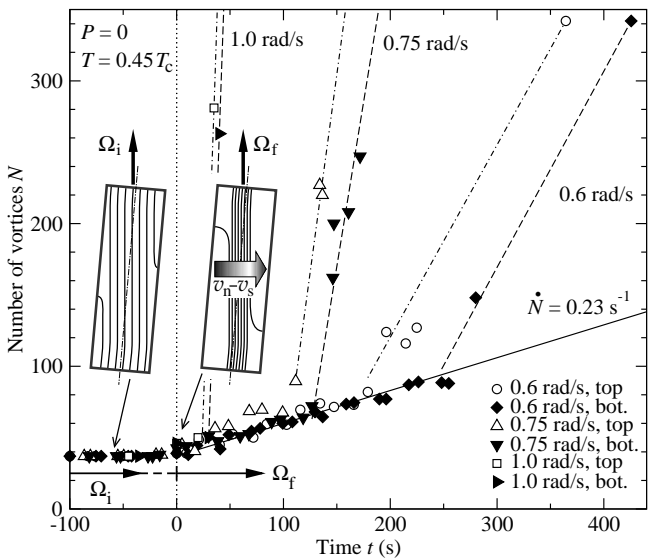}}
\caption{Vortex formation with a slow precursor which suddenly
develops to rapid turbulence. The number of vortices $N(t)$ is
recorded at the top and bottom of the sample. The solid line is
the average slow rate ${\dot N}$ while the dashed lines denote the
rapid turbulence. Initially the sample is in the equilibrium
vortex state at $\Omega_{\rm i}= 0.05\,$rad/s with $N \approx 37$
vortices and a few vortices connecting to the cylinder wall {\it
(left insert)}. Rotation is then increased to a new stable value
$\Omega_{\rm f}$, which is reached at $t = 0$. Three runs with
different $\Omega_{\rm f}$ are shown. During the ramp to
$\Omega_{\rm f}$, $N \approx {\rm const}$ while the flow builds up
and compresses the vortices in a central cluster {\it (right
insert)}. The precursor is attributed to an instability of the
single vortices, which extend across the counterflow region and
end at the cylindrical boundary. } \label{SlowVorFormation}
\end{figure}
%%%%%%%%%%%%%%%%%%%%%%%%%%%%%%%%%%%%%%%%%%%%%%%%%%%%%%%%

To reveal the precursor, we reduce the applied flow velocity, {\it
ie.} the value of $\Omega_{\rm f}$. The result is shown in
Fig.~\ref{SlowVorFormation}, where we plot the growth in the
vortex number $N(t)$ with time. The new feature is the slow
approximately linear increase in $N$ (solid line), before rapid
turbulence sets in (dashed lines).  At $\Omega_{\rm f} =
0.6\,$rad/s the slow increase lasts for more than 200\,s,
generating approximately 1 vortex in 5\,s until about 20\% of the
maximum number of vortices have been created. At larger
$\Omega_{\rm f}$ the slow increase is shorter in duration, {\it
eg.} at $\Omega_{\rm f} = 1\,$rad/s rapid turbulence starts
already after 30\,s. The slow vortex generation at roughly
constant rate $\dot N$ we identify as the precursor.

Two further observations about the precursor follow from
Fig.~\ref{SlowVorFormation}: (1) Vortex formation proceeds
independently in different parts of the sample. At $\Omega_{\rm f}
= 0.6\,$rad/s it takes more than 300\,s for a vortex created at
one end of the sample to reach the other end \cite{Experiment}.
Still, vortex formation at the top and bottom is observed to
proceed at roughly the same rate. Thus the precursor is not
localized (as would be {\it eg.} a vortex mill).

(2) In Fig.~\ref{SlowVorFormation} a second method has been used
to introduce vortices in flow, to start from a more controlled
initial vortex configuration. The equilibrium vortex state at a
low rotation of $\Omega_{\rm i} = 0.05\,$rad/s is employed as
starting point. Here a few vortices in the outer peripheral ring
next to the cylinder wall are not rectilinear but connect to the
wall (since the sample and rotation axes can only be aligned to
within about one degree, see inserts in
Fig.~\ref{SlowVorFormation}). To appreciate the influence of these
curved vortices, the experiment was repeated differently.

A cluster with only rectilinear vortices, but with less than the
equilibrium number, can be prepared at higher temperatures and can
then be cooled to $T < 0.6\,T_{\rm c}$. As long as this cluster is
separated by a sufficiently wide counterflow annulus from the
cylindrical boundary, $\Omega$ can be increased or decreased
without change in $N$ at any temperature down to our lowest value
of $0.35\,T_{\rm c}$. If $\Omega$ is reduced too much, the cluster
makes contact with the cylindrical boundary, some outermost
vortices become curved, and during a subsequent increase of
$\Omega$, while $T < 0.5\,T_{\rm c}$, the behavior in
Fig.~\ref{SlowVorFormation} is reproduced. Therefore to observe
vortex multiplication at $\Omega_{\rm f}$, we conclude that a
curved vortex connecting to the cylindrical wall is required. The
number of these initially curved vortices increases with
$\Omega_{\rm i}$ and thus also the rate $\dot N$ increases with
$\Omega_{\rm i}$. Nevertheless, the behavior is also observed when
$\Omega_{\rm i} = 0$ and one starts from a remnant vortex.

At a low value of $\Omega_{\rm f}$ a curved vortex which connects
to the cylindrical boundary spends a long time expanding axially
along the sample. The vortex segment adjacent to the wall moves in
the counterflow created by the rotation and reorients partially
along the flow owing to its self-induced velocity. It is thus
expected to become unstable with respect to the formation of
Kelvin waves \cite{Donnelly,Glaberson}. The expanding waves may
then reconnect with the wall and generate new separated loops.
This interpretation for the linear precursor in
Fig.~\ref{SlowVorFormation} explains qualitatively its distinct
features such as its abrupt temperature dependence or its low
threshold velocity. To analyze the precursor in more detail we
examine vortex dynamics numerically.

%%%%%%%%%%%%%%%%%%%%%%%%%%%%%%%%%%%%%%%%%%%%%%%%%%%%%%%%%%%%%%%%%%%
\begin{figure}[tb]
\centerline{\includegraphics[width=\linewidth]{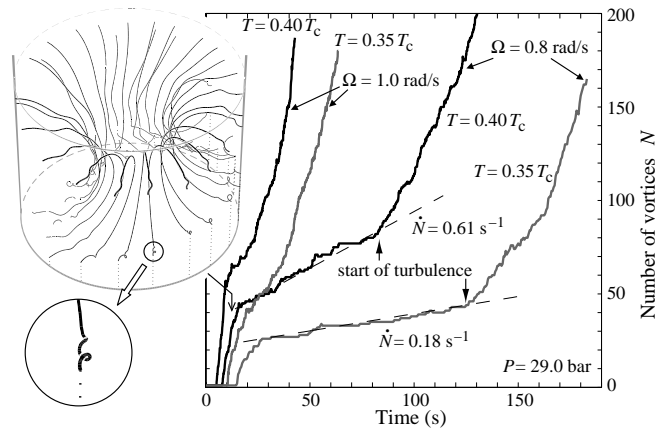}}
\caption{ Simulation of the measurements in
Fig.~\ref{SlowVorFormation}. The total number of separate vortices
$N(t)$ {\it vs.} time is followed in a cylinder of radius 3\,mm
and length 10\,mm. The calculation is started from a vortex ring
of 2\,mm radius in the azimuthal plane. The ring is unstable in
azimuthal flow and generates via the Kelvin wave instability
\cite{Glaberson,Donnelly} tens of vortices in a rapid burst, which
gives the configuration shown in the {\it insert}. After this
initial burst slow vortex formation at constant average rate
${\dot N}$ starts. Here each new vortex is produced from
Kelvin-waves expanding on an isolated vortex (as shown in the {\it
insert}) which is blown up to ring-like shape and then reconnects
at the boundary. The later rapid growth in vortex number is
dominated by inter-vortex interactions. }\label{SimSlowVorForm}
\end{figure}
%%%%%%%%%%%%%%%%%%%%%%%%%%%%%%%%%%%%%%%%%%%%%%%%%%%%%%%%%%%%%%%%%%%

%%%%%%%%%%%%%%%%%%%%%%%%%%%%%%%%%%%%%%%%%%%%%%%%%%%%%%%%%%%%%%%%%%%
\begin{figure}[t]
\centerline{\includegraphics[width=0.95\linewidth]{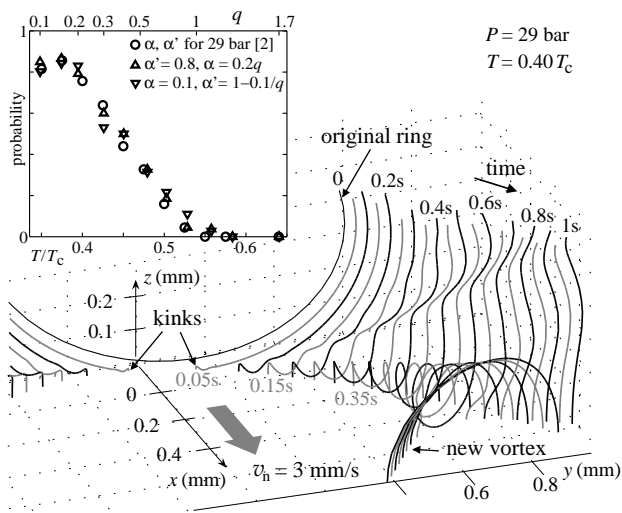}}
\caption{As a model of the precursor, the collision of a vortex
ring with a plane wall (at $z=0$) is calculated. A ring of 0.5\,mm
radius is initially above the wall, with the plane of the ring
tilted by $\pi/10$ from the $x=0$ plane. In the frame of reference
of this figure, uniform flow of the normal component at $v_{\rm n}
= 3\,$mm/s is applied in the $x$ direction. The different contours
show the ring at 0.05\,s intervals.  The sharp kinks at the wall
reconnections induce Kelvin waves on the loop. These are able to
grow in the applied flow only on one of the two legs formed in the
reconnection (here on the right). The largest wave reconnects with
the wall and a new loop is then separated. The conditions for the
growth of Kelvin waves are fulfilled in such collisions only in
the under-damped temperature regime: The {\it insert} on the top
shows the probability that a new vortex is created when the
original ring is initially placed at $z=1\,$mm with random
orientation. This probability depends on the ratio
$q=\alpha/(1-\alpha')$, shown on the top axis, rather than on
$\alpha$ or $\alpha'$ separately. } \label{WallReconnection}
\end{figure}
%%%%%%%%%%%%%%%%%%%%%%%%%%%%%%%%%%%%%%%%%%%%%%%%%%%%%%%%%%%%%%%%%%%

{\it Simulation:}---The velocity $\bm{v}_{\rm L}$ of a vortex line
element is determined from \cite{Donnelly}
\begin{equation}
\bm{v}_{\rm L}=\bm{v}_{\rm s} +\alpha \hat{\bm{s}} \times
(\bm{v}_{\rm n}-\bm{v}_{\rm s}) -\alpha' \hat{\bm{s}} \times
[\hat{\bm{s}} \times (\bm{v}_{\rm n}-\bm{v}_{\rm s})]\,.
\label{vl}
\end{equation}
Here $\hat{\bm{s}}$ is a unit vector parallel to the vortex line element
and $\alpha$ and $\alpha'$ are the mutual friction parameters measured in
Ref.\ \cite{Bevan}. From the form of this equation we can expect that the
solutions for $\bm{v}_{\rm L}$ can be classified by the parameter
$q=\alpha/(1-\alpha')$. We calculate the evolution of the vortex
configuration from Eq.~\eqref{vl}, with $\bm{v}_{\rm s}$ obtained from the
Biot-Savart law and the boundary conditions derived from an additional
solution of the Laplace equation \cite{Simulation,idealbnd}. We apply
background flow of $\bm{v}_{\rm n}$ in different geometries, including
rotating flow in a cylinder and flow in a pipe with a uniform or parabolic
velocity profile.

We find that at low vortex density new vortices are generated only
from expanding Kelvin waves which intersect with a wall. The
growth of these waves depends strongly on the orientation of the
vortex segment with respect to the flow \cite{Glaberson}, but it
starts from reconnection kinks \cite{Svistunov}. It is these sharp
kinks which prove to be essential in the simulations for continued
generation of new vortices \cite{Reconnection}. The kinks are
primarily produced when an expanding Kelvin wave hits the wall. In
a rotating cylinder this early stage is marked by roughly linear
growth in $N$, as seen in Fig.~\ref{SimSlowVorForm}.  Its duration
in time is similar to the measurements in
Fig.~\ref{SlowVorFormation} and it also ends in rapid turbulence.
The model of this single-vortex instability at the wall is studied
in Fig.~\ref{WallReconnection}. Here we see the expansion of
Kelvin waves from a reconnection kink at the wall and a later
reconnection again with the wall, which produces one new loop. The
probability of this process has been calculated in the insert. It
displays a similar abrupt temperature dependence as measured in
Fig.~\ref{SlowVorFormTempDependence}.

In numerical calculations self-sustained growth of $N$ is not
started as readily as in our experiment. Uniform flow along a
plane wall in Fig.~\ref{WallReconnection} does not support
continuous growth. In rotating flow the initial configuration in
Fig.~\ref{SimSlowVorForm} had to be specially engineered since
experimentally relevant configurations do not necessarily start a
continuous process in simulations. Clearly the explanation of this
difference is an interesting physical question. Self-sustained
vortex multiplication has been previously demonstrated in highly
inhomogeneous bulk flow of the normal component \cite{Samuels}.
When walls are present we find that the inhomogeneity of the
applied flow is not essential to maintain continuous growth in
$N$.  Nevertheless, the flow geometry affects the probability to
start self-sustained vortex generation from a single seed loop,
which is, for instance, larger in a circular pipe than in a
rotating cylinder (at the same value of $q$).

{\it Conclusions:}---Our results demonstrate that in the
under-damped regime of vortex motion intrinsic vortex formation
has to be described in an ideal superfluid as a sequence of
multiple processes. It starts with the nucleation of the first
vortex, is followed by surface-mediated multiplication of more
vortex loops, and it finally goes over in rapid turbulence when
inter-vortex interactions become possible. These individual steps
have been difficult to separate in earlier work. This is because
in superfluid $^4$He with strong surface pinning the solid
surfaces are covered with a plentiful source of remanent vortices
which easily start rapid turbulence when flow is applied. A second
reason is that experimentally oscillating flow has been easiest to
achieve and there the accumulation of a vortex tangle occurs
differently, as was recently reported from measurements with a
vibrating grid \cite{Lancaster}. We have here focused on the
missing link in this chain, the slow surface-mediated precursor to
rapid turbulence. If no other mechanism of vortex multiplication
intervenes, as is the case in steady flow of $^3$He-B with clean
smooth surfaces, then this process will take over.


\begin{thebibliography}{99}
\bibitem{idealbnd} As there is no firm evidence of pinning
  \cite{Parts}, we use the model of an ideal surface. With surface friction
  or weak pinning, the single-vortex transition moves to higher
  $q$.
\bibitem{Reconnection} The reconnection distance is a fixed parameter. Changing it by an order of magnitude
causes no qualitative changes.
\end{thebibliography}
\end{document}